\documentclass[sigconf,screen]{acmart}

\usepackage{booktabs} % For formal tables
\usepackage{flushend}
\usepackage[ruled]{algorithm2e} % For algorithms
\usepackage{multirow}
\usepackage{subfigure}
\usepackage{threeparttable}
\usepackage{latexsym}
\usepackage{amsmath}
\usepackage{enumitem}
\usepackage{color}
\usepackage{xcolor,colortbl}
\usepackage{tcolorbox}
\newcommand{\tabincell}[2]{\begin{tabular}{@{}#1@{}}#2\end{tabular}}

\newcommand{\para}[1]{\smallskip\noindent{\bf {#1}. }}
\newcommand{\emoji}[1]{\includegraphics[width=1em]{emoji_images/#1.png}}
\usepackage{tikz}
\newcommand*{\circled}[1]{\lower.7ex\hbox{\tikz\draw (0pt, 0pt)%
    circle (.5em) node {\makebox[1em][c]{\small #1}};}}

\newcommand{\approach}{SEntiMoji\xspace}

\makeatletter
\def\subsubsection{\@startsection{subsubsection}{3}%
  \z@{.5\linespacing\@plus.7\linespacing}{.1\linespacing}%
  {\normalfont\itshape}}
\makeatother

\setcopyright{acmcopyright}
\acmPrice{15.00}
\acmDOI{10.1145/3338906.3338977}
\acmYear{2019}
\copyrightyear{2019}
\acmISBN{978-1-4503-5572-8/19/08}
\acmConference[ESEC/FSE '19]{Proceedings of the 27th ACM Joint European Software Engineering Conference and Symposium on the Foundations of Software Engineering}{August 26--30, 2019}{Tallinn, Estonia}
\acmBooktitle{Proceedings of the 27th ACM Joint European Software Engineering Conference and Symposium on the Foundations of Software Engineering (ESEC/FSE '19), August 26--30, 2019, Tallinn, Estonia}

\begin{document}

\title{\approach: An Emoji-Powered Learning Approach for Sentiment Analysis in Software Engineering}

% \author{Zhenpeng Chen, Yanbin Cao, Xuan Lu}
% \affiliation{%
%   \institution{Key Lab of High-Confidence Software Technology, MoE (Peking University)}
%   \city{Beijing}
%   \country{China}
% }
% \email{{czp,caoyanbin,luxuan}@pku.edu.cn}

\author{Zhenpeng Chen}
\affiliation{%
  \institution{Key Lab of High-Confidence Software Technology, MoE (Peking University)}
  \city{Beijing}
  \country{China}
}
\email{czp@pku.edu.cn}

\author{Yanbin Cao}
\affiliation{%
  \institution{Key Lab of High-Confidence Software Technology, MoE (Peking University)}
  \city{Beijing}
  \country{China}
}
\email{caoyanbin@pku.edu.cn}

\author{Xuan Lu}
\affiliation{%
  \institution{Key Lab of High-Confidence Software Technology, MoE (Peking University)}
  \city{Beijing}
  \country{China}
}
\email{luxuan@pku.edu.cn}

\author{Qiaozhu Mei}
\affiliation{%
  \institution{School of Information, University of Michigan}
  \city{Ann Arbor}
  \country{USA}
}
\email{qmei@umich.edu}

\author{Xuanzhe Liu}\authornote{Corresponding author: Xuanzhe Liu (xzl@pku.edu.cn).}
\affiliation{%
  \institution{Key Lab of High-Confidence Software Technology, MoE (Peking University)}
  \city{Beijing}
  \country{China}
}
\email{xzl@pku.edu.cn}

%\renewcommand{\shortauthors}{Z. Chen et al.}

\begin{abstract}
Sentiment analysis has various application scenarios in software engineering (SE), such as detecting developers' emotions in commit messages and identifying their opinions on Q\&A forums. However, commonly used out-of-the-box sentiment analysis tools cannot obtain reliable results on SE tasks and the misunderstanding of technical jargon is demonstrated to be the main reason. Then, researchers have to utilize labeled SE-related texts to customize sentiment analysis for SE tasks via a variety of algorithms. However, the scarce labeled data can cover only very limited expressions and thus cannot guarantee the analysis quality. To address such a problem, we turn to the easily available emoji usage data for help. More specifically, we employ emotional emojis as noisy labels of sentiments and propose a representation learning approach that uses both Tweets and GitHub posts containing emojis to learn sentiment-aware representations for SE-related texts. These emoji-labeled posts can not only supply the technical jargon, but also incorporate more general sentiment patterns shared across domains. They as well as labeled data are used to learn the final sentiment classifier. Compared to the existing sentiment analysis methods used in SE, the proposed approach can achieve significant improvement on representative benchmark datasets. By further contrast experiments, we find that the Tweets make a key contribution to the power of our approach. This finding informs future research not to unilaterally pursue the domain-specific resource, but try to transform knowledge from the open domain through ubiquitous signals such as emojis.
\end{abstract}

\begin{CCSXML}
<ccs2012>
<concept>
<concept_id>10002951.10003317.10003347.10003353</concept_id>
<concept_desc>Information systems~Sentiment analysis</concept_desc>
<concept_significance>500</concept_significance>
</concept>
</ccs2012>
\end{CCSXML}

\ccsdesc[500]{Information systems~Sentiment analysis}

\keywords{Emoji; Sentiment analysis; Software engineering}

\maketitle

\section{Introduction}\label{intro}
% Affective states can play an important role in various aspects of collaborative work, including productivity, task quality, group rapport, user focus, etc.~\cite{cscwChoudhuryC13}. As a typical kind of collaborative activity, 
Software development is a highly collaborative activity that is susceptible to the affective states of developers~\cite{OrtuADTMT15,msrMurgiaTAO14,GuzmanB13,jssJuradoM15}. On the one hand, negative sentiments can make developers underperform in the software projects that they contribute to~\cite{GarciaZS13} and lead to longer issue fixing time~\cite{OrtuADTMT15}. Therefore, project managers need to stay aware of the affective states of developers, in order to detect negative sentiments and take timely  necessary actions to ensure high productivity of developers~\cite{GuzmanB13}. On the other hand, understanding the sentiments in developers' discussions can help software practitioners be aware of collective opinions about specific SE topics (e.g., adding a new feature in software) or artifacts (e.g., software libraries or APIs) ~\cite{reGuzmanM14,D0008ZBPLO18,msrPleteaVS14,scamRahmanRK15,icse0008ZBPL19}, which can then help their further actions about these topics and their usage and improvement of these artifacts.
% Therefore, project managers need to stay aware of the affective states of developers, in order to uncover negative sentiments and take necessary actions immediately~\cite{GuzmanB13}. In addition, understanding the sentiments of developers in forum discussions can help assess the quality of different software artifacts such as APIs and provide advice for the usage of them during software development~\cite{ZhangH13,D0008ZBPLO18}.

Sentiment analysis has been a frequently used natural language processing (NLP) technique in SE. It aims to identify the affective states and subjective opinions in texts. Many out-of-the-box sentiment analysis tools (e.g., SentiStrength~\cite{ThelwallBPCK10}) not designed for SE-related texts have been applied to SE tasks, but recent work has indicated that they cannot produce reliable results on SE tasks~\cite{JongelingSDS17}. Furthermore, Islam and Zibran~\cite{IslamZ17} applied SentiStrength to an SE-related dataset and found that misunderstanding of domain-specific meanings of words (namely \textit{technical jargon} in the rest of this paper) accounts for the most misclassifications. Such a finding inspires a series of research efforts in recent years to generate SE-specific datasets and develop customized sentiment analysis methods based on them~\cite{AhmedBIR17,ese/CalefatoLMN18,D0008ZBPLO18}. Various machine-learning or deep-learning techniques have been applied, but the performance still falls short of producing adequate results in practice for some SE tasks~\cite{D0008ZBPLO18}.

% Unfortunately, traditional ways to capture the developers' sentiments, such as face-to-face meetings, interviews, and surveys~\cite{hsiWrobel13}, are not feasible and efficient any more, in terms of the large volume of current SE projects where developers are often  geographically distributed and communicate through textual channels such as technical Q\&A sites and issue tracking systems. To make up the deficiency of these traditional approaches, researchers in SE community turn to off-the-shelf sentiment analysis tools (e.g., SentiStrength~\cite{ThelwallBPCK10}) for help~\cite{GuzmanAL14,GuzmanB13,reGuzmanM14,SinhaLS16,GarciaZS13,msrPleteaVS14,Rousinopoulos17,scamRahmanRK15}. However, these tools are trained on non-technical texts and demonstrated to be unable to obtain satisfactory results on SE tasks~\cite{JongelingSDS17}. Islam and Zibran applied SentiStrength to a SE-specific dataset and found that misunderstanding of domain-specific meanings of words (namely \textit{technical jargon} in the rest of this paper) accounts for the most misclassifications~\cite{IslamZ17}. Such a finding inspires a series of research efforts in recent years to annotate SE-specific labeled datasets and develop customized sentiment analysis methods based on them~\cite{AhmedBIR17,ese/CalefatoLMN18,D0008ZBPLO18}. 
% However, although various machine-learning or deep-learning techniques are applied, the performance still falls far short of identifying sentiment expressions in SE-related discussions ~\cite{D0008ZBPLO18}.

One possible reason behind the poor performance could be the customized methods that are trained based on scarce SE-related labeled data (only thousands of samples), and inevitably lack the knowledge of other expressions that are not contained in them. Given the large volume of English vocabulary, these missing expressions are indeed non-trivial. To tackle this problem, a straightforward solution is to annotate abundant SE-related texts with sentiment labels. However, manual annotation on a large scale is quite difficult, time-consuming, and error-prone~\cite{FelboMSRL17}. Instead, recent work in NLP attempted to employ emotional emojis as noisy labels of sentiments on social media~\cite{FelboMSRL17}. As emojis become an emerging ubiquitous language used worldwide~\cite{Lu16}, emoji-labeled texts are easily available, which can help tackle the scarcity of manually labeled texts and thus benefit the sentiment analysis tasks~\cite{FelboMSRL17}. Inspired by this work, we aim to explore the emoji usage data into sentiment analysis in SE. Here then comes a question. Where should we extract such data?

In fact, emojis not only pervasively exist in social media, but are also widely adopted in the communication of developers to express sentiment~\cite{githubemoji}. For example, in the post  ``\emph{thanks for writing this great plugin\emoji{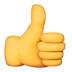}}''\footnote{\url{https://github.com/MikaAK/s3-plugin-webpack/issues/65}, retrieved in November 2018.} on GitHub, the emoji ``\emoji{519}'' can be considered a positive sentiment signal.
In order to ensure the representativeness of emoji usage data, we employ posts containing emojis from both Twitter (a typical social media platform) and GitHub (a typical software development platform). Here, the core insight is: \textit{posts from GitHub can provide more technical information beyond the limited labeled data, while posts from Twitter can help learn more general sentimental patterns that are shared in both technical and non-technical communication}.

We propose \emph{\approach}, an emoji-powered learning approach for sentiment analysis in SE. Through \approach, vector representations of texts are first derived based on modeling how emojis are used alongside words on Twitter and GitHub. These sentiment-aware representations are then used to predict the sentiment polarities on the labeled data. 
To evaluate the performance of \approach, we  answer two research questions: 

   $\bullet$ \noindent \textbf{RQ1}: \emph{How does \approach perform compared to the existing sentiment analysis methods in SE?}  By rigorous contrast experiments, we find that \approach can significantly outperform existing sentiment analysis methods in SE on all the selected benchmark datasets. This finding indicates that the incorporation of emoji usage data is a promising solution to the SE-customized sentiment analysis.
    
   $\bullet$ \noindent \textbf{RQ2}: \emph{Which training corpora contribute more to the power of \approach? } We find that GitHub posts do not make a key contribution. The combination of large-scale Tweets and a small amount of labeled data can already achieve satisfactory performance. Such results highlight the significance of the general sentimental patterns learned from Tweets. Our finding informs future research not to focus only on the domain-specific resource. Instead, they can also pay attention to the general sentimental patterns that can be easily extracted from the open domain.
   
%   that the researchers should not focus on only the technical jargon and do not have to make too many efforts on labeling data for the sentiment analysis in SE task.

The main contributions of this paper are as follows:
\begin{itemize}[leftmargin=*]
% \item To the best of our knowledge, we are the first to leverage emojis as the instrument in sentiment analysis for SE tasks.
\item We propose an emoji-powered learning approach for SE-customized sentiment analysis, which utilizes Tweets to capture the general sentimental expressions and GitHub posts as well as manually labeled data to incorporate the technical jargon.
\item We demonstrate the effectiveness of \approach for SE tasks using four representative benchmark datasets. \approach can significantly improve the state-of-the-art results on all these datasets.
\item We explore the underlying reasons behind the performance of \approach by rigorous contrast experiments and provide future research some insightful implications.
\end{itemize}

The rest of this paper is organized as follows. Section~\ref{related} summarizes the literature related to this study. Section~\ref{methodology} presents the workflow of \approach. Section~\ref{evaluation} compares \approach with baseline methods on representative benchmark datasets and answers the two research questions based on the achieved results. Section~\ref{lesson} summarizes the lessons learned in this study and the implications. Section~\ref{limitation} discusses the threats that could affect the validity of this study, followed by concluding remarks in Section~\ref{conclusion}.

\section{Related Work} \label{related}
We start with the literature related to this study. Our research is particularly inspired by two streams of literature: sentiment analysis in SE and emojis in sentiment analysis.

\subsection{Sentiment Analysis in SE}
In recent years, sentiment analysis has been widely applied in SE for enhancing software development, maintenance, and evolution~\cite{PanichellaSGVCG15,msrBlazB16,JongelingDS15, OrtuADTMT15, GuzmanB13,mantyla2016mining,ortu2016arsonists,hsiWrobel13,wrobel2016towards,GachechiladzeLN17,msrSouzaS17}.
% Analyzing the sentiment on SE platforms can help understand the affective states and attitudes of developers~\cite{msrMurgiaTAO14,GarciaZS13,casconTouraniJA14}. On the one hand, the affective states can affect the creativity and productivity of developers in collaborative projects~\cite{cscwChoudhuryC13,GarciaZS13,OrtuADTMT15}. For example, Ortu \textit{et al.} found the affective states are linked with the issue fixing time on issue tracking systems~\cite{OrtuADTMT15}. Positive sentiments and emotions such as joy can reduce the resolution time, while negative sentiments such as sadness can have the opposite effect~\cite{OrtuADTMT15}. Therefore, if negative emotions can be detected through automatic approaches, typical actions can be taken in time to help developers defuse these situations so as to improve their performance. On the other hand, mining sentiment in the discussions about specific SE topics (e.g., adding a new feature in the software) or SE artifacts (e.g., software libraries or APIs) can help developers be aware of collective opinions about them~\cite{reGuzmanM14,D0008ZBPLO18,msrPleteaVS14,scamRahmanRK15}, which can then benefit their further actions about these topics and their usage and improvement of these artifacts. 
Most of these studies used the out-of-the-box sentiment analysis tools (e.g., SentiStrength~\cite{ThelwallBPCK10}, NLTK~\cite{aclBirdL04}, and Stanford NLP~\cite{ManningSBFBM14}) trained on non-technical texts. Among these tools, SentiStrength is considered to be the most widely adopted one in SE studies~\cite{GuzmanAL14,ChowdhuryH16a,GuzmanB13,reGuzmanM14,JongelingDS15,OrtuADTMT15,SinhaLS16,TouraniA16,NovielliCL15,GarciaZS13}. However, some researchers noticed unreliable results when directly employing such tools for SE tasks~\cite{JongelingSDS17,D0008ZBPLO18}. Jongeling \textit{et al.}~\cite{JongelingSDS17} observed the disagreement among these existing tools on the datasets in SE and found that the results of several SE studies involving these sentiment analysis tools cannot be confirmed when a different tool is used. To investigate the challenges in sentiment analysis in SE, Islam and Zibran~\cite{IslamZ17} applied the most popular SentiStrength to some labeled issue comments extracted from JIRA issue tracking system and conducted an in-depth qualitative study to uncover twelve difficulties in identifying the sentiments of SE-related texts by analyzing the misclassified samples. Among the identified difficulties, lacking domain-specific knowledge is demonstrated to be the most dominant, accounting for about 81\% of the classification errors. Since then, how to effectively leverage SE-related texts to introduce technical jargon becomes the main direction of the sentiment analysis in SE~\cite{AhmedBIR17,ese/CalefatoLMN18,D0008ZBPLO18}. Against such a background, many SE-customized sentiment analysis tools and methods are proposed, including SentiStrength-SE~\cite{IslamZ17}, Senti-CR~\cite{AhmedBIR17}, Senti4SD~\cite{ese/CalefatoLMN18}, etc. We take them along with the most popular SentiStrength as baseline methods in this study and introduce them detailedly in Section~\ref{baseline}.

\subsection{Emojis in Sentiment Analysis}
%Since the debut on Twitter and Instagram, emojis quickly expanded territory to various platforms, of course also including some SE communities such as GitHub~\cite{githubreport,githubemoji}. Many research efforts have been devoted to studying their prevalence~\cite{zhenpeng18,Lu16,LjubesicF16}. As an important factor to their popularity, the non-verbal functions of emojis are also well studied~\cite{CramerJT16,HuGSNL17}. Among various functions, expressing sentiment is demonstrated to be the most popular intention~\cite{HuGSNL17} and this characteristic sheds light on current sentiment analysis. 
Traditional sentiment analysis in NLP is mainly performed in \emph{unsupervised} or \emph{supervised} ways. Unsupervised tools (e.g., SentiStrength) simply make use of lists of words annotated with sentiment polarity to determine the overall sentiment of a given text. However, fixed word lists cannot cope with the dynamic nature of the natural language~\cite{GiachanouC16}. Then, researchers started to use labeled text to train sentiment classifiers for different purposes in a supervised way. However, it is time-consuming to manually annotate text on a large scale, thus resulting in a scarcity of labeled text. To tackle this problem, many researchers attempted to perform sentiment analysis in a \emph{distantly supervised} way. For example, they used binarized emoticons~\cite{LiuLG12} and specific hashtags~\cite{colingDavidovTR10} as a proxy for the emotional contents of a text. Recent studies extended the distant supervison to emojis, a more diverse set of noisy labels~\cite{FelboMSRL17,elsa}. As emojis are becoming increasingly popular~\cite{zhenpeng18,Lu16,AiLLW0M17} and have the ability to express emotions~\cite{HuGSNL17}, they are considered benign noisy labels of sentiments in current sentiment analysis~\cite{FelboMSRL17,elsa}. The sentiment information contained in the emoji usage data can supplement the limited manually labeled data.
%For example, they used the emotional non-verbal cues (such as emoticons or emojis) as the proxy of sentiment labels and represented the texts co-used with the same sentiment proxy similarly through deep learning algorithms~\cite{LiuLG12,ZhaoDWX12,FelboMSRL17}. The informative and sentimental representations were then used for further training of the sentiment classifier. During this process, the sentiment information from the texts co-used with non-verbal cues supplement the scarce labeled data. As emojis are more popular than other non-verbal cues such as emoticons and have the ability to convey various emotions, they are considered as benign proxy of sentiments in current sentiment analysis~\cite{ermes,FelboMSRL17}.

Recently, to address the challenge of sentiment analysis in SE, researchers also started to analyze emoticons and emojis in software development platforms so as to find some potential solutions. Claes \textit{et al.}~\cite{ClaesMF18} investigated the use of emoticons in open source software development. Lu \textit{et al.}~\cite{githubemoji} analyzed the emoji usage on GitHub and found that emojis are often used to express sentiment on this platform. Furthermore, Imtiaz \textit{et al.}~\cite{ImtiazMCRBM19} directly used emojis as the indicators of developers' sentiments on GitHub. Calefato \textit{et al.}~\cite{ese/CalefatoLMN18} and Ding \textit{et al.}~\cite{Ding00L18} took emoticons into account in their proposed sentiment analysis techniques built on SE-related texts. All of them demonstrated the feasibility of leveraging these emotional cues to benefit sentiment analysis in SE. Following this line of research, this study leverages the large-scale emoji usage from both technical and open domains to address sentiment analysis in SE.
\section{methodology}\label{methodology}
As we mentioned before, sentiment analysis is a traditional NLP task. In this section, we give a brief description to the fundamental concepts of related techniques and then illustrate the workflow of \approach detailedly.

\subsection{Preliminaries}
We first present some background knowledge on the NLP techniques that will be used in this paper, including word embedding, Long Short-Term Memory (LSTM) network, and fine-tuning.

\subsubsection{Word Embedding}
To eliminate the discrete nature of words, word embedding is employed by NLP tasks to encode every single word into a continuous vector space as a high dimensional vector. It is usually trained by learning from large scale corpus via GloVe~\cite{PenningtonSM14}, CBOW~\cite{Mikolov2013Efficient}, or skip-gram algorithm~\cite{Mikolov2013Efficient}. In this paper, the skip-gram algorithm is employed for word embedding. This algorithm scans each example in the training corpus and uses each word it has scanned as an input to predict words within a certain range before and after this word. By such a prediction task, words that commonly occur in a similar context are embedded closely in the vector space, which captures the semantic relationship between words. 

\subsubsection{LSTM}
Recurrent neural network (RNN)~\cite{Rumelhart1988Learning} is a kind of neural network specialized for processing sequential data such as texts. It connects computational units of the network in a directed cycle such that at each time step, a unit in RNN takes both the current input and the hidden state of the same unit from the previous time step as the input. Due to the recurrent nature of RNN, it can capture the sequential information, which is important to NLP tasks. However, due to the well-known gradient vanishing problem, vanilla RNNs are difficult to train to capture long-term dependency for sequential texts. LSTM~\cite{Hochreiter1997Long} addresses this problem by introducing a gating mechanism to determine when and how the states of hidden layers can be updated. Each LSTM unit contains a memory cell, an input gate, a forget gate, and an output gate. The input gate controls the input activations into the memory cell, and the output gate controls the output flow of cell activations into the rest of the network. The memory cells in LSTM store the sequential states of the network, and each memory cell has a self-loop whose weight is controlled by the forget gate. The LSTM structure ensures that the gradient of the long-term dependencies cannot vanish. 

\subsubsection{Fine-tuning}
Labeled data are often limited for NLP tasks, especially new ones. For such tasks, training a neural network from scratch with limited data may result in over-fitting. One approach to getting around this problem is to take a network model, which has been trained for a given task, to perform the target task. This process is commonly called fine-tuning~\cite{YosinskiCBL14}. Assuming the target task is similar to the original task, fine-tuning enables us to take advantages of the prior efforts on feature extraction (i.e., the pre-trained parameters of the network). The first step of fine-tuning is to replace the last fully-connected layer of the original network with a new one that can output a probability vector whose dimension is the desired number of classes in the target task. Then, we can use the labeled data for the target task to fine-tune parameters of the original network and make the network be suitable for the new task.

\subsection{The \approach Approach}
\approach is an SE-customized sentiment classifier trained based on a small amount of labeled SE-related data as well as large scale emoji-labeled data from both Twitter and GitHub. As emojis are widely used to express sentiment~\cite{HuGSNL17,githubemoji}, we learn sentiment-aware representations of texts by using emoji prediction as an instrument. More specifically, we use emojis as noisy labels of sentiments and learn vector representations of sentences by predicting which emojis are used in a sentence. Sentences that tend to occur with the same emoji are represented similarly, which captures the sentiment relationship between sentences and can thus benefit the downstream sentiment classification. Since such a representation model trained on large scale Tweets has been off-the-shelf, i.e., \emph{DeepMoji} model~\cite{FelboMSRL17,deepmoji}. We directly build \approach upon DeepMoji. It takes a two-stage approach: 1) fine-tune DeepMoji using emoji-labeled texts from GitHub to incorporate technical knowledge. The fine-tuned model is still a representation model based on the emoji-prediction task and we call it \emph{DeepMoji-SE}; 2) use DeepMoji-SE to obtain vector representations of the sentiment-labeled texts and then use these vectors as features to train the sentiment classifier.

Next, we describe the existing DeepMoji model and the two-stage learning process in details.

\subsubsection{DeepMoji Model}
Felbo \textit{et al.}~\cite{FelboMSRL17} learned DeepMoji model through predicting emojis used in Tweets. To this end, they collected 56.6 billion Tweets (denoted as $T$), selected the top 64 emojis in this corpus, and excluded the Tweets that do not contain any of these emojis. For each remaining Tweet, they created separate samples for each unique emoji in it. For example, ``\emph{Good idea!}\emoji{519}\emoji{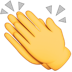}\emoji{521}'' can be separated into two samples, i.e., (``\emph{Good idea!}'', \emoji{519}) and (``\emph{Good idea!}'', \emoji{521}). Finally, the generated 1.2 billion samples (denoted as $ET$) were used to train the representation model. 

\begin{figure}
\includegraphics[width=0.75\columnwidth]{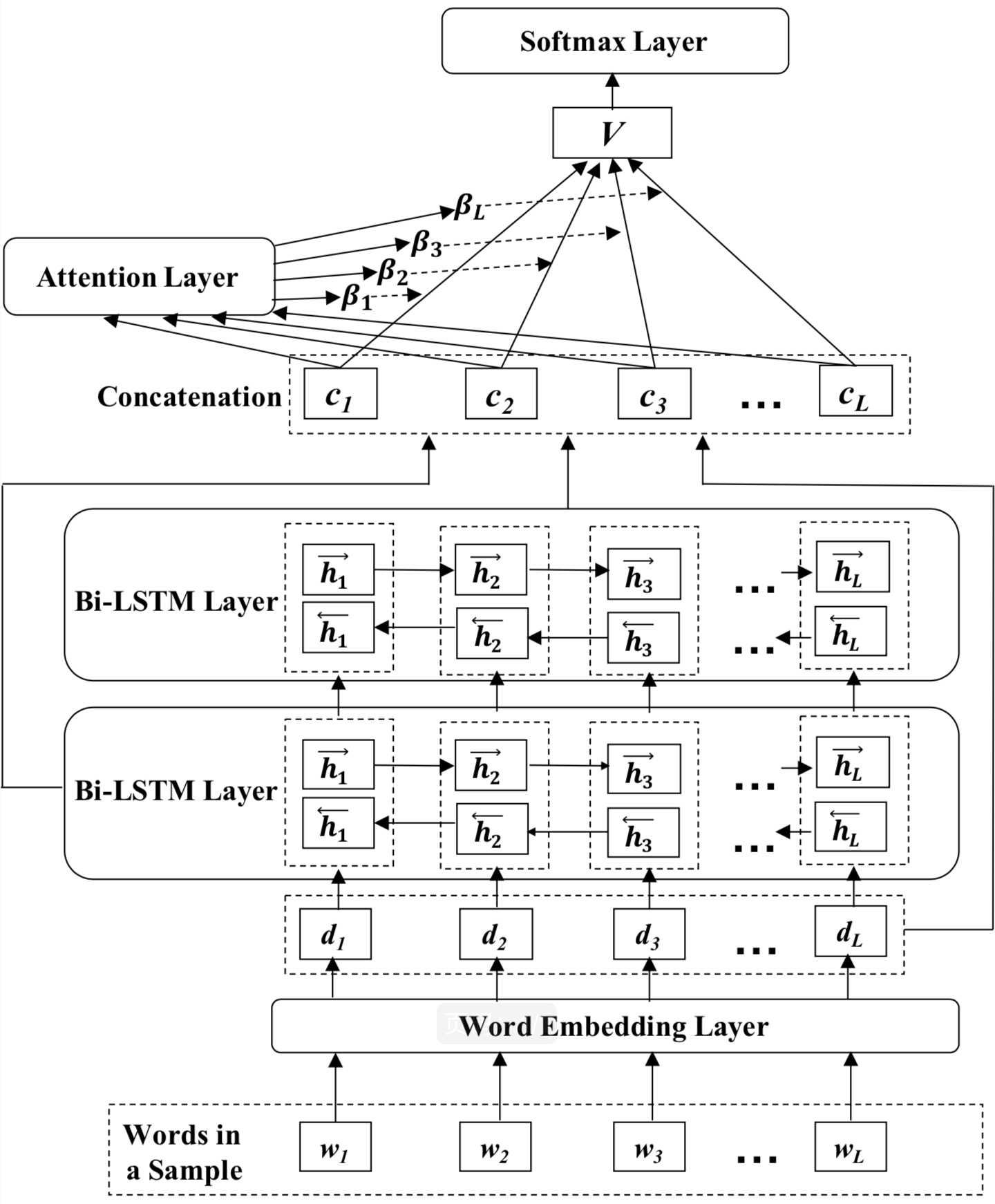}
\caption{The architecture of DeepMoji.}\label{fig:emoji_predict}
\end{figure}

The model architecture is illustrated in Figure~\ref{fig:emoji_predict}. First, for a given sample, words in it are inputted into the word embedding layer pre-trained on $T$. In this step, each word can be represented as a unique vector. These word vectors are then processed by two bi-directional LSTM layers and one attention layer. Through these steps, the sample can be represented as one sentence vector instead of several word vectors. Finally, the softmax layer treats the sentence vector as the input and outputs the probabilities that this sample contains each emoji. Taking the real emoji contained in each sample in $ET$ as ground truth, the model learns parameters by minimizing the output error of the softmax layer. The details of the model architecture are described below.

% \para{Word Embedding Layer} The word embedding layer is pre-trained based on $T$ and can project each word into a continuous vector space, where words that usually occur in similar contexts are represented close to each other~\cite{Mikolov2013Efficient}. Through this layer, each sample in $ET$ can be denoted as $(x, e)$, where $x = [d_1, d_2, ..., d_L]$ as the word vector sequences of the plain text removed emoji ($d_i$ as the vector representation of the $i$-th word) and $e$ as the emoji contained in the sample.

\para{Word Embedding Layer} The word embedding layer is pre-trained based on $T$. Through this layer, each sample in $ET$ can be denoted as $(x, e)$, where $x = [d_1, d_2, ..., d_L]$ as the word vector sequences of the plain text removed emoji ($d_i$ as the vector representation of the $i$-th word) and $e$ as the emoji contained in the sample.

% \para{Bi-Directional LSTM Layer} The Long Short-Term Memory network (LSTM) is suitable for processing the sequential texts due to its recurrent nature~\cite{Hochreiter1997Long}. Words in one sample can be processed by the LSTM orderly. At each time step, one word is inputted into the layer and then the layer combines the current input word and the knowledge from the past time steps (i.e., the past words) to update the layer state. In this way, LSTM can capture the long semantic dependencies in the sequential texts rather than considering each word individually. Compared the traditional LSTM, at one time step, bi-directional LSTM can take both the past and the future words (i.e., the contexts) of the current word into consideration, which is demonstrated to be more effective in some natural language processing tasks such as text classification~\cite{ZhouQZXBX16}. 
\para{Bi-Directional LSTM Layer} Given the input $x = [d_1, d_2, ..., d_L]$, at step $t$, LSTM computes unit states of the network as follows:

\begin{subequations}
\begin{equation*}
i^{(t)} = \sigma(U_id_t+W_ih^{(t-1)}+b_i),
\end{equation*}
\begin{equation*}
f^{(t)}=\sigma(U_fd_t+W_fh^{(t-1)}+b_f),
\end{equation*}
\begin{equation*}
o^{(t)}=\sigma(U_od_t+W_oh^{(t-1)}+b_o),
\end{equation*}
\begin{equation*}
c^{(t)}=f_t \odot c^{(t-1)} + i^{(t)} \odot tanh(U_cd_t+W_ch^{(t-1)}+b_c),
\end{equation*}
\begin{equation*}
h^{(t)} = o^{(t)} \odot tanh(c^{(t)}),
\end{equation*}
\end{subequations}

where $i^{(t)}$, $f^{(t)}$, $o^{(t)}$, $c^{(t)}$, and $h^{(t)}$ denote the states of the input gate, forget gate, output gate, memory cell, and hidden layer at step $t$. $W$, $U$, $b$, and $\odot$ denote the recurrent weights, input weights, biases, and element-wise product, respectively. In order to take both the past and the future words of the current word at each time step into consideration, we employ bi-directional LSTMs instead of the traditional LSTM. Each bi-directional LSTM network contains two sub-networks (i.e., a forward network and a backward network) to encode the sequential contexts of each word in the two directions respectively. We thus compute an encoded vector $h_i$ of each word vector $d_i$ by concatenating the latent vectors from both directions:

% \begin{subequations}
% \begin{equation*}
% \mathop{h_i}\limits ^{\rightarrow}=\mathop{LSTM}\limits ^{\longrightarrow}(d_i),
% \end{equation*}
% \begin{equation*}
% \mathop{h_i}\limits ^{\leftarrow}=\mathop{LSTM}\limits ^{\longleftarrow}(d_i),
% \end{equation*}
\begin{equation*}
h_i = \mathop{h_i}\limits ^{\rightarrow} ||  \mathop{h_i}\limits ^{\leftarrow},
\end{equation*}
% \end{subequations}

where $\mathop{h_i}\limits ^{\rightarrow}$ and $\mathop{h_i}\limits ^{\leftarrow}$ denote the forward and backward states of $d_i$, respectively. 

In order to enable the unimpeded information flow in the whole model, the outputs of the two LSTM layers and the word embedding layer are concatenated by the skip-connection algorithm~\cite{Hermans2013Training}, then as input into the attention layer. Specifically, each word of the input sample is further represented as $c_i$:

\begin{equation*}
c_i = d_i || h_{i1} || h_{i2},
\end{equation*}

where $d_i$, $h_{i1}$, and $h_{i2}$ represent the encoded vectors of the $i$-th word extracted from the word embedding layer and the first and second bi-directional LSTM layers. 

\para{Attention Layer} Since not all words contribute equally to the overall sentiment polarity of the sample, the model employs the attention mechanism~\cite{VaswaniSPUJGKP17} to determine the importance of each word. The attention score (i.e., the importance) of the $i$-th word is computed as:

\begin{equation*}
\alpha_i = \frac{exp(Wc_i)}{\sum_{j=1}^Lexp(Wc_j)},
\end{equation*}

where $W$ is the weight matrix of the attention layer. Then the sample can be represented as the weighted sum of all words in it:

\begin{equation*}
V=\sum_{i=1}^L\alpha_ic_i.
\end{equation*}

\para{Softmax Layer} The final representation $V$ is inputted into the softmax layer to output a 64-dimension probability vector, each element of which denotes the probability that this sample contains one specific emoji. 

DeepMoji learns parameters by minimizing the cross entropy between the output probability vectors and the one-hot representations of the emoji actually contained in each sample. Through such a learning phase, the plain texts occur with the same emoji can be represented similarly. 
% If we input a sentence into DeepMoji, we can extract the final representation $V$ to represent it. 

\subsubsection{Stage 1: Fine-tuning DeepMoji Using GitHub Data}
As DeepMoji is trained on Tweets, we need some developer-generated emoji-labeled texts to customize it into an SE-specific setting. To this end, we use the conversation data (i.e., issues, issue comments, pull requests, and pull request comments) in the \emph{GitHub-Emoji dataset} collected by Lu \textit{et al.}~\cite{githubemoji}, which cover more than one hundred million posts on GitHub, to fine-tune the parameters of DeepMoji. 

We first conduct the following procedures to pre-process the GitHub data. We tokenize all the texts into words and convert all the words into lowercase. Then we replace URLs, references, code snippets, and mentions with specific tokens in case the concrete contents of these texts influence the sentiment polarity. For example, we replace any code snippets with ``[code].'' From the processed GitHub data, we extract the posts containing emojis and then create separate samples for each unique emoji in each post. Then the samples containing any of the 64 emojis predicted by DeepMoji are finally selected for fine-tuning DeepMoji. We denote about 1 million remaining samples as $EG$.

% contain about 95 million posts, from which we extract the ones containing any of the 64 emojis predicted by DeepMoji. The extracted posts account for 69\% of the total emoji usage in the raw corpus. Following the procedure of DeepMoji, we create separate samples for each unique emoji in each selected post. We denote the about 1 million generated samples as $EG$.

Next, we use $EG$ to fine-tune DeepMoji by the chain-thaw approach~\cite{FelboMSRL17}. The chain-thaw approach sequentially unfreezes and fine-tunes a single layer in DeepMoji.
By training each layer separately, it can adjust the individual patterns across the network with a reduced risk of over-fitting. 
More specifically, it first fine-tunes the softmax layer, and then fine-tunes each layer individually starting from the first layer (i.e., the word embedding layer) in the network. Finally, the entire model is trained with all layers and we obtain an SE-customized representation model that learns from emoji-labeled Tweets and GitHub posts. We refer to it as \emph{DeepMoji-SE} in this paper.

% For each sample in $EG$, we use the plain text as the input and the emoji contained in it as the ground-truth to fine-tune the original hyperparameters of DeepMoji and thus obtain DeepMoji-SE, a new SE-customized representation model.

\subsubsection{Stage 2: Training the Sentiment Classifier}
Based on DeepMoji-SE, labeled SE-related data (denoted as $LD$) can be represented as sentiment-aware vectors, which are further used as features to learn the final sentiment classifier.
% Based on DeepMoji-SE, we use the manually labeled SE-related data (denoted as $LD$) to learn the final sentiment classifier. The learning phase is also a process of fine-tuning.

Specifically, the training phase of the sentiment classifier is also a process of fine-tuning. With other settings of the whole DeepMoji-SE unchanged, we replace its 64-dimension softmax layer with an $n$-dimension softmax layer, where $n$ denotes the number of sentiment polarities in the classification task. Then, we use $LD$ to fine-tune the parameters of DeepMoji-SE by the aforementioned chain-thaw approach.

% For each sample in $LD$, we input the text into the revised model and can generate a sentiment-aware representation $V$. Then $V$ is inputted into the $n$-dimension softmax layer to calculate the probability vector, each element of which represents the probability the text belongs to one specific sentiment polarity. Finally, the model re-learns hyperparameters by minimizing the divergence between the probability vector and the one-hot representation of the real sentiment label.

\section{Evaluation}\label{evaluation}
We then evaluate the performance of \approach by comparing it with some baseline methods on existing benchmark datasets.

\subsection{Baseline Methods}\label{baseline}
We employ four existing sentiment analysis methods for comparisons, including \emph{SentiStrength}, \emph{SentiStrength-SE}, \emph{SentiCR}, and \emph{Senti4SD}. SentiStrength is not designed for technical texts but has been the most popular sentiment analysis tool in previous SE studies, while the other three tools are specifically proposed for SE-related texts in recent years. We now describe these methods briefly:

\textbf{SentiStrength}~\cite{ThelwallBPCK10,sentistrength_tool} is a lexicon-based sentiment classifier trained for informal English texts rather than technical texts. It utilizes a dictionary of several word and phrase lists to compute the sentiment of texts. The dictionary contains a sentiment word strength list, a booster word list, a negating word list, an emoticon list, etc. For an input text, SentiStrength outputs a positive score and a negative score based on its coverage of the built-in dictionary. 
%The positive score ranges from 1 (not positive) to 5 (extremely positive), while the negative score ranges from -5 (not negative) to -1 (extremely negative).
Based on the algebraic sum of the two scores, SentiStrength can report a trinary score, i.e., 1 (positive), 0 (neutral), or -1 (negative).

\textbf{SentiStrength-SE}~\cite{IslamZ17,sentistrengthse} is a lexicon-based tool adapted from SentiStrength. It is developed based on the results obtained by running SentiStrength on  Group-1 of the JIRA dataset (an SE-related dataset that will be described in Section~\ref{benchmark}). It first identifies the reasons behind the misclassified cases and then makes efforts to address them. For example, it adapts the raw inherent dictionary of SentiStrength to contain some SE-related terms. 
%SentiStrength-SE is demonstrated to be superior to SentiStrength in terms of their performance on the Group-2 and Group-3 of JIRA dataset. 

\textbf{SentiCR}~\cite{AhmedBIR17,senticr} is a supervised sentiment analysis method originally proposed for code reviews. It computes Term Frequency - Inverse Document Frequency (TF-IDF)~\cite{aizawa2003information} of bag-of-words in a corpus as features and uses traditional machine learning algorithms to train the sentiment classifier. In addition, it applies SMOTE~\cite{ChawlaBHK02} for over-sampling to address class imbalance in the training data. In this study, we use the Gradient Boosting Tree~\cite{friedman2002stochastic} to reproduce this approach as recommended by its authors.
%Finally, it uses eight machine-learning algorithms to train the classifier and the Gradient Boosting Tree (GBT)~\cite{friedman2002stochastic} performs the best.

\textbf{Senti4SD}~\cite{ese/CalefatoLMN18,senti4sd} is a supervised sentiment analysis method for developer-generated texts. 
It leverages three kinds of features for sentiment classification, including lexicon-based features (based on the sentimental word list of SentiStrength), keyword-based features (such as uni-grams and bi-grams), and semantic features (based on the word embeddings trained on large scale posts collected from Stack Overflow). Finally, it uses Support Vector Machine~\cite{suykens1999least} to train the sentiment classifier.

Besides the four existing methods, we also adopt six variants of \approach as baseline methods, so as to measure the contribution of Tweets, GitHub posts, and manually labeled data to the overall performance of \approach. The six variants are described as follows:

\textbf{\approach-G} adopts the model architecture of DeepMoji and uses GitHub posts to directly train the DeepMoji-SE, instead of using GitHub data to fine-tune DeepMoji. Then the re-trained DeepMoji-SE is used for training the final sentiment classifier. Compared to \approach, \approach-G is trained purely based on GitHub data while not leveraging any Tweet.

\textbf{\approach-T} uses DeepMoji rather than DeepMoji-SE for the final training of the sentiment classifier in the second stage. Compared to \approach, \approach-T does not use any GitHub post. It can learn the knowledge of SE-related jargon only from the manually labeled data in the second stage.

\textbf{T-80\%}, \textbf{T-60\%}, \textbf{T-40\%}, and \textbf{T-20\%} are adapted from \approach-T. They differ from \approach-T only in the size of the labeled data. They randomly select 80\%, 60\%, 40\%, and 20\% of the labeled data used by \approach-T to train the sentiment classifier and keep other settings unchanged.

\subsection{Benchmark Datasets}\label{benchmark}
We compare the performance of our approach and the ten baseline methods on four representative datasets covering SE-related texts from different platforms, i.e., \emph{JIRA dataset}, \emph{Stack Overflow dataset}, \emph{Code Review dataset}, and \emph{Java Library dataset}. All of them are collected and annotated for sentiment analysis in SE and have been released online. More details of the four datasets are provided in the following:

\textbf{JIRA dataset}~\cite{OrtuMDTTMA16,jiradata} originally contained 5,992 issue comments extracted from the JIRA issue tracking system. It was divided into Group-1, Group-2, and Group-3, containing 392, 1,600, and 4,000 issue comments, respectively. As SentiStrength-SE is developed by analyzing Group-1 data, in order to compare these methods fairly, we exclude Group-1 comments from this dataset. Group-2 and Group-3 comments were labeled with love, joy, surprise, anger, sadness, fear, or neutral. In accordance with the previous study~\cite{NovielliGL08}, we consider love and joy as positive, sadness and fear as negative, and discard the surprise label to avoid introducing noise as it can match either positive or negative. For Group-2 comments, the original annotations by all three coders were released. We assign each comment with positive, negative, or neutral if it was annotated with the corresponding label by at least two coders. Under such criteria, comments that cannot match any sentiment label are excluded. For each comment in Group-3, authors of the JIRA dataset directly released its golden emotion label, which was assigned if at least two raters marked the presence of this emotion~\cite{OrtuADTMT15}. After excluding the surprise labels, we also discard the comments with no label or opposite sentiment labels. Finally, we have 2,573 remained comments from Group-2 and Group-3, 42.9\% of which are positive, 27.3\% neutral, and 29.8\% negative.

\textbf{Stack Overflow dataset}~\cite{ese/CalefatoLMN18,senti4sd} contains 4,423 samples and covers four types of Stack Overflow posts, including questions, answers, question comments, and answer comments. The raw dataset was originally extracted from the Stack Overflow dump from July 2008 to September 2015. Then, to ensure a balanced polarity distribution, 4,800 posts were remained based on their sentiment detected by SentiStrength. Next, each post was labeled by three coders with positive, negative, or neutral. Based on the annotations, the posts annotated with opposite polarity labels were excluded by its authors. Then the final label of each text was determined via majority voting. Finally, among the remaining 4,423 posts, 34.5\% are positive, 38.3\% neutral, and the rest 27.2\% negative.

\textbf{Code Review dataset}~\cite{AhmedBIR17,senticr} contains 1,600 code review comments extracted from code review repositories of 20 popular open source software projects. The dataset originally contained 2,000 comments, each of which was independently labeled by three coders with positive, negative, or neutral. For the comments where raters had different opinions, their final labels were determined through discussion. Then the distribution of the labeled review comments were: 7.7\% positive, 19.9\% negative, and 72.4\% neutral. Due to the serious class imbalance, its authors randomly excluded a subset of the majority class (i.e., neutral) and aggregated the remaining neutral and positive comments into ``non-negative'' class. Finally, among the remaining 1,600 comments, 24.9\% are negative and 75.1\% non-negative.

\begin{table}[!tp]
\centering
\footnotesize
\caption{Benchmark datasets used in this study.}
\label{data}
\begin{tabular}{lrrrr}
\hline
\multirow{2}{*}{Dataset} & \multirow{2}{*}{Size} & \multicolumn{3}{c}{Polarity distribution} \\
\cline{3-5}
& &  Positive & Neutral & Negative \\
\hline
JIRA & 2,573 &  1,104 (42.9\%)&  702 (27.3\%) &  767 (29.8\%)\\
%\hline
Stack Overflow & 4,423 &  1.527 (34.5\%)&  1,694 (38.3\%) &  1,202 (27.2\%) \\
%\hline
Code Review & 1,600 & \multicolumn{2}{c}{1,202 (75.1\%)} &  398 (24.9\%)\\
%\hline
Java Library & 1,500& 131 (8.7\%) & 1,191 (79.4\%) & 178 (11.9\%)\\
\hline
\end{tabular}
\end{table}

\textbf{Java Library dataset}~\cite{D0008ZBPLO18,stackoverflow} contains 1,500 sentences about Java libraries or Java APIs extracted from the Stack Overflow dump of July 2017. Each sentence was annotated by two coders. The coders labeled each sentence with a sentiment score from -2 to 2 (-2 indicates strong negative, -1 weak negative, 0 neutral, 1 weak positive, and 2 strong positive). After the labeling process, two of the authors worked on conflict resolution and gave each sentence a consistent and double-checked sentiment label. Finally, among the 1,500 collected sentences, 8.7\% are positive, 79.4\% neutral, and 11.9\% negative.

We summarize the statistics of the four datasets in Table~\ref{data}.

\begin{table*}[!tp]
\begin{threeparttable}
\scriptsize
\centering
\caption{Performance of our emoji-powered models and existing sentiment analysis tools.}
\label{performance}
\begin{tabular}{c|cr|c||ccc||c||cc||cccc}
\hline
Dataset & Class & Metric & SentiStrength & SentiStrength-SE & SentiCR & Senti4SD & \approach & \approach-G & \approach-T & T-80\%  & T-60\%  & T-40\%  & T-20\%\\
\hline
\hline
\multirow{10}{*}{JIRA} & \multirow{3}{*}{Pos} & Precision & 0.847 & 0.936 & 0.950 & 0.880 & 0.947 & 0.938 & 0.939 & \cellcolor{blue!20}\textbf{0.952} & 0.948 & 0.940 & 0.950\\
& & Recall &  0.889 & 0.922 & 0.919 & 0.921 & \cellcolor{blue!20} \textbf{0.945} & 0.920 & 0.933 &  0.931 & 0.935 & 0.937 & 0.921\\
& & F-score & 0.868 & 0.929 & 0.934 & 0.900 & \cellcolor{blue!20}\textbf{0.946} & 0.928 & 0.936 & 0.941 & 0.941 & 0.938 & 0.935\\
\cline{2-14}
& \multirow{3}{*}{Neu} & Precision & 0.614 & 0.710 & 0.735 & 0.741 & \cellcolor{blue!20}\textbf{0.823} & 0.780 & 0.808 & 0.795 & 0.798 & 0.779 & 0.775\\
& & Recall & 0.634 & 0.844 & \cellcolor{blue!20}\textbf{0.904} & 0.731 & 0.880 & 0.869 & 0.866 & 0.887 & 0.883 & 0.861 & 0.861\\
& & F-score & 0.623 & 0.771 & 0.811 & 0.736 & \cellcolor{blue!20}\textbf{0.850} & 0.822 & 0.845 & 0.838 & 0.838 & 0.817 & 0.815\\
\cline{2-14}
& \multirow{3}{*}{Neg} & Precision &  0.775 & 0.871 & \cellcolor{blue!20}\textbf{0.929} & 0.835 & 0.922 & 0.888 & 0.913 & 0.905 & 0.905 & 0.901 & 0.879\\
& & Recall & 0.699 & 0.734 & 0.768 & 0.789 & \cellcolor{blue!20}\textbf{0.864} & 0.818 & 0.839 & 0.835 & 0.835 & 0.813 & 0.825\\
& & F-score & 0.735 & 0.796 & 0.840 & 0.811 & \cellcolor{blue!20}\textbf{0.892} & 0.851 & 0.874 & 0.868 & 0.868 & 0.854 & 0.851\\
\cline{2-14}
& \multicolumn{2}{c|}{Accuracy} & 0.763 & 0.846 & 0.872 & 0.830 & \cellcolor{blue!20}\textbf{0.904} & 0.876 & 0.893 & 0.891 & 0.891 & 0.880 & 0.876\\
\hline
\hline

\multirow{10}{*}{\tabincell{c}{Stack\\ Overflow}} & \multirow{3}{*}{Pos} & Precision & 0.887 & 0.908 & 0.868 & 0.904 & \cellcolor{blue!20}\textbf{0.932} & 0.908 & 0.921 & 0.926 & 0.922 & 0.923 & 0.918\\
& & Recall & 0.927 & 0.823 & 0.921 & 0.915 & \cellcolor{blue!20}\textbf{0.940} & 0.864 & 0.937 & 0.929 & 0.930 & 0.922 & 0.912\\
& & F-score & 0.907 & 0.863 & 0.894 & 0.910 & \cellcolor{blue!20}\textbf{0.936} & 0.885 & 0.929 & 0.928 & 0.926 & 0.926 & 0.915\\
\cline{2-14}
& \multirow{3}{*}{Neu} & Precision & \cellcolor{blue!20}\textbf{0.922} & 0.726 & 0.783 & 0.829 & 0.840 & 0.753 & 0.841 & 0.830 & 0.826 & 0.819 & 0.813\\
& & Recall & 0.632 & 0.784 & 0.838 & 0.772 & \cellcolor{blue!20}\textbf{0.842} & 0.805 & 0.840 & 0.837 & 0.830 & 0.836 & 0.828\\
& & F-score & 0.750 & 0.754 & 0.809 & 0.800 & \cellcolor{blue!20}\textbf{0.841} & 0.778 & \cellcolor{blue!20}\textbf{0.841} & 0.834 & 0.828 & 0.827 & 0.820\\
\cline{2-14}
& \multirow{3}{*}{Neg} & Precision & 0.674 & 0.755 & 0.843 & 0.778 & \cellcolor{blue!20}\textbf{0.846} & 0.762 & 0.854 & 0.840 & 0.837 & 0.834 & 0.825 \\
& & Recall & \cellcolor{blue!20}\textbf{0.931} & 0.759 & 0.686 & 0.841 & 0.833 & 0.736 & 0.838 & 0.827 & 0.827 & 0.810 & 0.811\\
& & F-score & 0.780 & 0.757 & 0.753 & 0.808 & 0.838 & 0.748 & \cellcolor{blue!20}\textbf{0.845} & 0.833 & 0.828 & 0.828 & 0.817\\
\cline{2-14}
& \multicolumn{2}{c|}{Accuracy} & 0.815 & 0.800 & 0.826 & 0.840 & \cellcolor{blue!20}\textbf{0.873} & 0.806 & \cellcolor{blue!20}\textbf{0.873} & 0.866 & 0.862 & 0.858 & 0.852\\
\hline
\hline

\multirow{7}{*}{\tabincell{c}{Code\\ Review}} & \multirow{3}{*}{Non-Neg} & Precision & 0.806 & 0.795 & \cellcolor{blue!20}\textbf{0.872} & 0.840 & 0.869 & 0.801 & 0.850 & 0.857 & 0.857 & 0.839 & 0.818\\
& & Recall & 0.814 & 0.919 & 0.895 & 0.912 & 0.941 & 0.943 & \cellcolor{blue!20}\textbf{0.958} & 0.950 & 0.948 & 0.957 & 0.948\\
& & F-score & 0.809 & 0.852 & 0.883 & 0.875 & \cellcolor{blue!20}\textbf{0.904} & 0.865 & 0.900 & 0.900 & 0.899 & 0.894 & 0.877\\
\cline{2-14}
& \multirow{3}{*}{Neg} & Precision & 0.506 & 0.537 & 0.660 & 0.638 & 0.762 & 0.633 & \cellcolor{blue!20}\textbf{0.793} & 0.778 & 0.774 & 0.778 & 0.717\\
& & Recall & 0.474 & 0.238 & \cellcolor{blue!20}\textbf{0.600} & 0.475 & 0.572 & 0.291 & 0.491 & 0.521 & 0.527 & 0.447 & 0.364\\
& & F-score & 0.488 & 0.372 & 0.627 & 0.544 & \cellcolor{blue!20}\textbf{0.653} & 0.393 & 0.603 & 0.614 & 0.618 & 0.561 & 0.469 \\
\cline{2-14}
& \multicolumn{2}{c|}{Accuracy} & 0.712 & 0.761 & 0.823 & 0.804 & \cellcolor{blue!20}\textbf{0.849} & 0.780 & 0.841 & 0.841 & 0.841 & 0.829 & 0.801\\
\hline
\hline

\multirow{10}{*}{\tabincell{c}{Java\\ Library}} & \multirow{3}{*}{Pos} & Precision & 0.202 & 0.320 & 0.553 & 0.472 & 0.849 & 0.737 & \cellcolor{blue!20}\textbf{0.878} & 0.825 & 0.803 & 0.828 & 0.728 \\
& & Recall & \cellcolor{blue!20}\textbf{0.369} & 0.224 & 0.318 & 0.203 & 0.329 & 0.230 & 0.281 & 0.214 & 0.222 & 0.192 & 0.142\\
& & F-score & 0.206 & 0.262 & 0.401 & 0.266 & \cellcolor{blue!20}\textbf{0.472} & 0.349 & 0.426 & 0.337 & 0.343 & 0.309 & 0.235\\
\cline{2-14}
& \multirow{3}{*}{Neu} & Precision & 0.858 & 0.824 & \cellcolor{blue!20}\textbf{0.883} & 0.860 & 0.880 & 0.834 & 0.869 & 0.863 & 0.862 & 0.850 & 0.836\\
& & Recall & 0.768 & 0.929 & 0.910 & 0.926 & 0.964 & 0.967 & 0.971 & 0.967 & 0.972 & 0.972 & \cellcolor{blue!20}\textbf{0.979}\\
& & F-score & 0.810 & 0.873 & 0.896 & 0.893 & \cellcolor{blue!20}\textbf{0.920} & 0.895 & 0.917 & 0.912 & 0.914 & 0.907 & 0.902\\
\cline{2-14}
& \multirow{3}{*}{Neg} & Precision & 0.396 & 0.487 & 0.546 & 0.522 & 0.729 & 0.489 & 0.734 & 0.737 & 0.735 & 0.706 & \cellcolor{blue!20}\textbf{0.769}\\
& & Recall & 0.434 & 0.183 & \cellcolor{blue!20}\textbf{0.593} & 0.463 & 0.583 & 0.223 & 0.513 & 0.518 & 0.490 & 0.412 & 0.335\\
& & F-score & 0.412 & 0.265 & 0.565 & 0.481 & \cellcolor{blue!20}\textbf{0.644} & 0.302 & 0.599 & 0.598 & 0.586 & 0.513 & 0.448\\
\cline{2-14}
& \multicolumn{2}{c|}{Accuracy} & 0.693 & 0.778 & 0.821 & 0.807 & \cellcolor{blue!20}\textbf{0.863} & 0.814 & 0.859 & 0.849 & 0.849 & 0.838 & 0.829\\
\hline
\end{tabular}
\begin{tablenotes}
\item \emph{Note}: For each metric, the highest value is highlighted with shading.
\end{tablenotes}
\end{threeparttable}
\end{table*}

% \subsection{Evaluation Metrics}
% In line with the recent study on assessing sentiment analysis on SE tasks~\cite{D0008ZBPLO18}, we measure the performance of each method in terms of the precision, recall, and F-score of each polarity class as well as the overall accuracy. 

% \textbf{Accuracy} measures how often one method makes the correct prediction and is defined as below:
% \begin{equation*}
% accuracy = \frac{\textrm{\#right predicted samples}}{\textrm{\#total samples}},
% \end{equation*}

% \textbf{Precision} represents the exactness of one method. The precision of a given polarity class \textit{c} is measured as:
% \begin{equation*}
% precision@c = \frac{\textrm{\#right predicted samples belonging to class \textit{c}}}{\textrm{\#total samples predicted as class \textit{c}}},
% \end{equation*}

% \textbf{Recall} represents the sensitivity of one method. The recall of a given polarity class \textit{c} is calculated as:
% \begin{equation*}
% recall@c = \frac{\textrm{\#right predicted samples belonging to class \textit{c}}}{\textrm{\#total samples belonging to class \textit{c}}},
% \end{equation*}

% \textbf{F-score} is a combination of precision and recall. The F-score of a given polarity class \textit{c} is computed as the harmonic mean of precision@c and recall@c:

% \begin{equation*}
% F-score@c = \frac{2*precision@\textit{c}*recall@\textit{c}}{precision@\textit{c}+recall@\textit{c}},
% \end{equation*}

\subsection{Experimental Setting}
In this study, we evaluate \approach and all the baseline methods on the four benchmark datasets. To make a fair comparison, for each dataset, we test each method in the same five-fold cross validation setting. More specifically, for a given dataset, we randomly split it into five equal subsets and thus can test each method for five times. Each time, we use one unique subset as the test set and the remaining four subsets as the training set to train all the methods. Since SentiStrength and SentiStrength-SE are not based on supervised machine learning but on a set of rules, we do not re-train them. 

In accordance with the recent study on assessing sentiment analysis methods on SE tasks~\cite{D0008ZBPLO18}, we measure the performance of each method in terms of the \emph{precision}, \emph{recall}, and \emph{F-score} of each polarity class as well as the overall \emph{accuracy}. For each dataset, we calculate the aforementioned metrics of each method for five turns in cross validation and finally report the mean value of each metric.

Because we adopt so many metrics to evaluate each method from different aspects, it is hard for us to conclude whether one method outperforms the others just based on the difference in one specific metric. To test whether the performance gap between methods is statistically significant, we apply the non-parametric McNemar's test~\cite{dietterich1998approximate}. This test suits well for our purpose as it does not require the normal distribution of data and is also adopted in a related study~\cite{jss/IslamZ18}. Through the five-fold cross validation, each method has output a predicted label for each sample. To compare method $A$ with method $B$ via McNemar's test, we need to derive the number of samples misclassified by $A$ but not by $B$ (denoted as $n_{01}$) and the number of samples misclassified by $B$ but not by $A$ (denoted as $n_{10}$). Then we can compute the statistic $\frac{{(|n_{01}-n_{10}|-1)}^2}{{n_{01}+n_{10}}}$, which is distributed as $\chi^2$ with 1 degree of freedom. Note that when we apply multi-hypothesis tests, we use Benjamini-Yekutieli procedure~\cite{Bjcorrection} for correction. More specifically, we use ``\emph{p.adjust}'' in R and set ``\emph{p.adjust.methods}'' as "\emph{BY}" to adjust the $p$-value of each individual test to obtain more strict and reliable results. The performance difference is considered statistically significant, only if the $p$-value of the computed statistic is lower than a pre-specified significance level.
\subsection{Research Questions and Results}\label{result}
In Table~\ref{performance}, we summarize the performance of \approach and baseline methods. For each combination of  dataset and metric, we highlight the best result with shading. Based on the achieved results, we first want to investigate whether \approach performs better than the other existing methods.

\subsubsection{RQ1: How does \approach perform compared to the existing sentiment analysis methods in SE?}

To answer this question, we compare \approach with four existing methods (i.e., SentiStrength, SentiStrength-SE, SentiCR, and Senti4SD). It is observed that \approach can achieve the best performance on most metrics. To verify whether the superiority of \approach is statistically significant, for each dataset, we perform McNemar's test between the results of \approach and each existing method, and show statistics in  Table~\ref{testresult}. Based on the results of statistical tests, we now derive a straightforward answer to \emph{RQ1}: \approach can significantly outperform these existing sentiment analysis methods on all of the datasets. Next, we want to compare the results more thoroughly.

%\begin{table}[h]
%\begin{threeparttable}
%\centering
%\footnotesize
%\caption{McNemar's Statistics between the results of \approach and other existing methods.}
%\label{testresult}
%\begin{tabular}{lrrrr}
%\hline
%Dataset & SentiStrength & SentiStrength-SE & SentiCR & Senti4SD\\
%\hline
%JIRA & 231.84** & 62.51** & 23.62** & 85.71**\\
%Stack Overflow & 95.39** & 128.55** & 76.96** & 38.69**\\
%Code Review & 102.20** & 55.84** & 7.32** & 19.71** \\
%Java Library & 160.16** & 69.52** & 20.56** & 40.79** \\
%\hline
%\end{tabular}
%\begin{tablenotes}
%\item \emph{Note}: ** means significant at 1\% level.
%\end{tablenotes}
%\end{threeparttable}
%\end{table}
\begin{table}[!tp]
\begin{threeparttable}
\centering
\footnotesize
\caption{McNemar's statistics between the results of \approach and other existing methods on each dataset, with p-values in parentheses.}
\label{testresult}
\begin{tabular}{lrrrr}
\hline
 Methods & JIRA & Stack Overflow & Code Review & Java Library \\
\hline
SentiStrength & 231.843** & 95.394** & 102.202** & 160.160**\\
 &(0.000)&(0.000)&(0.000)&(0.000)\\
SentiStrength-SE & 62.510**  & 128.548** & 55.840** & 69.522**\\
 &(0.000)&(0.000)&(0.000)&(0.000)\\
SentiCR & 23.616** & 76.963** & 7.320* & 20.556** \\
 &(0.000)&(0.000)&(0.023)&(0.000)\\
Senti4SD & 85.708** & 38.686** & 19.711** & 40.786** \\
 &(0.000)&(0.000)&(0.000)&(0.000)\\
\hline
\end{tabular}
\begin{tablenotes}
\item \emph{Note}: Results labeled with ** / * are significant at the 1\% / 5\% level.
\end{tablenotes}
\end{threeparttable}
\end{table}

We first compare \approach with the most widely used SentiStrength, which is an out-of-the-box sentiment analysis tool without SE-customized efforts. In terms of overall accuracy, \approach can outperform SentiStrength by at least 0.1 on the JIRA, Code Review, and Java Library datasets. 
By comparison, the performance difference on the Stack Overflow dataset is only 0.058. This ``outlier'' can be attributed to the creation process of the Stack Overflow dataset. Calefato \textit{et al.}~\cite{ese/CalefatoLMN18} created this dataset by sampling the originally collected posts based on their sentiment scores computed by SentiStrength. It is easier for SentiStrength to correctly classify the samples selected by itself, which thus results in a relatively small performance gap between \approach and SentiStrength on the Stack Overflow dataset.

Then we want to compare \approach with the SE-customized methods (i.e., SentiStrength-SE, SentiCR, and Senti4SD). In general, SentiCR performs the best among the three existing methods as it can achieve the highest accuracy and the highest F-score of each polarity class on each dataset except the Stack Overflow dataset. On the Stack Overflow dataset, it performs slightly worse than Senti4SD. It is reasonable as the semantic features used by Senti4SD are extracted based on the embeddings trained on large scale Stack Overflow corpus and thus Senti4SD is more knowledgable than SentiCR when dealing with Stack Overflow posts. As SentiCR has an obvious advantage over other SE-customized tools in general, we then just compare our \approach with it. At a glance of the results, compared to SentiCR,
%SentiCR can outperform \approach in only several metrics with a small advantage. The mean value of the performance gap in the six metrics where SentiCR prevails is only 0.011. By comparison, 
\approach has an obvious superiority in most metrics. For example, on the JIRA dataset, the accuracy of \approach and SentiCR is 0.904 and 0.872, respectively. In other words, their error rates are respectively 0.096 and 0.128. The 0.032 difference in accuracy means \approach can correctly classify 3.2\% of the total samples more than SentiCR. The seemingly trivial difference is non-negligible. In terms of error rates, \approach can reduce 25\% of the samples misclassified by SentiCR on this dataset. In addition, the superiority is also evidenced by the results of statistical tests we mentioned before.

What's more, when carefully inspecting the results, we find the performance gap is particularly large in some  cases. For example, in terms of precision, the extent to which \approach outperforms SentiCR on the Java Library dataset is obviously larger than on other datasets. This phenomenon can be attributed to the sampling methods of different datasets. As the JIRA dataset is labeled with various emotion labels, we need to map the multi-class emotions into the trinary polarities and some samples are filtered due to ambiguity. With regard to the Stack Overflow and Code Review datasets, we find both of them are manually pre-processed and filtered to follow a not that skewed sentiment distribution during their creation process~\cite{ese/CalefatoLMN18,AhmedBIR17}. The sampling of these datasets makes the classification tasks considerably easier, so SentiCR and \approach do not show such obvious performance difference on these datasets. Compared to the other datasets, the Java Library dataset has a more imbalanced class distribution and thus the classification task on it is more like ``finding needles in a haystack.'' In such a situation, \approach can still achieve 0.296 and 0.183 higher precision over positive and negative samples while keeping the same level of recall, which further demonstrates the superiority of \approach.

% \finding{RQ1 main finding: \approach can significantly outperform the existing sentiment analysis methods on SE tasks, although  performance differences vary due to different creation processes of the benchmark datasets.}

Since \approach can achieve significant improvement on the benchmark datasets, we then want to investigate the reasons behind its power, which can provide insights for the further research.

\subsubsection{RQ2: Which training corpora contribute more to the power of \approach?}

To answer this question, we first compare \approach with \approach-G and \approach-T, to measure the contributions of Tweets and GitHub posts. The three methods share the same model architecture and use the same labeled data for training. They differ only in the external data used in representation learning. \approach uses Tweets to supply the general sentimental expressions and uses GitHub posts to introduce more technical jargon into the model. By comparison, besides the manually labeled data, \approach-G uses only GitHub posts, while \approach-T uses only Tweets.

We perform the McNemar's test to compare their performance. As shown in Table~\ref{TorG}, \approach-G performs significantly worse than \approach, while the performance difference between \approach and \approach-T is not statistically significant. Then we turn to Table~\ref{performance} for a more detailed comparison. \approach-T exactly has comparable results with \approach, with only 0.006 drop in average accuracy on the four datasets. It can even achieve higher values than \approach in several metrics. By comparison, the average decrease of \approach-G in accuracy is 0.053, which is 8.83 times that of \approach-T. In addition, \approach-G loses the advantage of our approach in the precision on the Java Library dataset. The precision@neg gained by \approach-G is only 0.489, which is even lower than the 0.546 achieved by SentiCR, let alone when compared with the precision levels of 0.729 and 0.734 gained by \approach and \approach-T. These findings reveal that Tweets contribute more than GitHub posts to the power of \approach. 
%In other words, as the manually labeled data can provide technical information, using social media data to capture more common expressions is more important than introducing more technical jargons via GitHub data.

\begin{table}[!tp]
\begin{threeparttable}
\footnotesize
\caption{McNemar's statistics between the results of \approach and its simplified versions on each dataset, with p-values in parentheses.}
\label{TorG}
\begin{tabular}{lrrrr}
\hline
Methods & JIRA & Stack Overflow & Code Review & Java Library \\
\hline
\approach-G & 19.066** &136.970** &  44.981** & 32.494**\\
 &(0.000)&(0.000)&(0.000)&(0.000)\\
\approach-T &  5.134 & 0.004 & 1.482 & 1.266\\
& (0.100) & (1.000) & (0.808) & (0.811)\\
\hline
\end{tabular}
\begin{tablenotes}
\item \emph{Note}: Results labeled with ** / * are significant at the 1\% / 5\% level.
\end{tablenotes}
\end{threeparttable}
\end{table}

\begin{table}[!tp]
\begin{threeparttable}
\footnotesize
\caption{McNemar's statistics between the results of \approach-T and its simplified versions on each dataset, with p-values in parentheses.}
\label{data_size}
\begin{tabular}{lrrrr}
\hline
Methods & JIRA & Stack Overflow & Code Review & Java Library \\
\hline
T-80\% & 0.114 & 4.571 & 0.000 & 2.086 \\
& (1.000) & (0.198) & (1.000) & (0.733)\\
T-60\% & 0.078 & 10.527* & 0.010 & 1.639\\
& (1.000) & (0.011) & (1.000) & (0.902)\\
T-40\% & 7.585* & 16.962** & 2.676 & 9.592*\\
& (0.041) & (0.000) & (0.552) & (0.015)\\
T-20\% & 9.551* & 32.761** & 29.184** & 17.204**\\
& (0.015)&(0.000)&(0.000)&(0.000)\\
\hline
\end{tabular}
\begin{tablenotes}
\item \emph{Note}: Results labeled with ** / * are significant at the 1\% / 5\% level.
\end{tablenotes}
\end{threeparttable}
\end{table}

% Does this finding mean the so-called technical jargon is not such important to our task? To answer this derivative question, 
Next, we measure the contribution of the labeled SE-related data. To this end, we compare \approach-T with T-80\%, T-60\%, T-40\%, and T-20\%, to verify whether the performance declines with the reduction of the labeled data. All of the these methods use only Tweets for representation learning, and then learn the technical knowledge only from the labeled data. The sole difference is the size of the labeled data used for training. 

Their performance differences from \approach-T can be observed in Table~\ref{performance}. In terms of overall accuracy, the decrease on the Code Review dataset is more obvious compared to on other datasets. When we use 20\% of it to train the model, the accuracy is 0.801, which is even lower than SentiCR with 100\% data. However, for the JIRA, Stack Overflow, Java Library datasets, when we scale down the labeled data, the accuracy gets just slightly worse. The accuracy under 20\% labeled data is 0.876, 0.852, and 0.829, respectively. These results are still higher than the performance of SentiCR trained on 100\% data. However, in terms of other metrics on specific datasets, we can observe some obvious decrease. For example, on the Java Library dataset, using 20\% labeled data for training can still achieve higher precision of positive and negative polarity compared to SentiCR, but it shows obviously lower recall levels.

To intuitively understand the changes in the performance when scaling down the labeled data, we illustrate McNemar's statistics between the results of \approach-T and the other four models with less labeled data in Table~\ref{data_size}. We find the change patterns vary from dataset to dataset. For the Stack Overflow dataset, the performance can be significantly worse with 80\% labeled data and the performance gap gets larger with the decrease of the labeled data. However, for the Code Review dataset, the performance maintains at a comparable level till 80\% of the labeled data are excluded. The findings reveal that the technical texts are essential in this task. However, the combination of Tweets and not so many technical texts is able to obtain satisfying results, which further demonstrate the importance of the general sentimental expressions incorporated by Tweets.

\subsection{Case Study}
To get a deeper insight into \approach, we perform a case study to investigate the examples misclassified by it and categorize some notable error classes in the following:
\begin{itemize}[leftmargin=*]
\item \textbf{Implicit sentiments.} The sentiments contained in some samples are exactly implicit. Take a negative post ``\emph{I would recommend that you check the documentation for what they do}'' as an example. The author is blaming someone that she/he should have checked the documentation already. However, this implication is so obscure that \approach can hardly capture. 

% is classified by \approach as negative. Actually, we ourselves also do not observe sentiment in such samples. These misclassifications may be attributed to the subjectivity of sentiment annotations. In some cases, raters were too sensitive in interpreting the samples and provided a positive or negative label even in presence of sentimental cues.
\item \textbf{Sentimental words in neutral sentences.} There are a lot of samples that contain obviously sentimental words but are neutral in fact. For example, in the neutral sample ``\emph{Because there are too many changes in Rails 3, without tests it may be very painful}'', a simply fact is being stated without negative or positive sentiments whereas the word ``\emph{painful}'' misleads \approach into classifying it as negative.
\item \textbf{Long samples.} Among misclassified instances, there are some long texts, such as ``\emph{I think the question is valid. I agree with the other responses, but it doesn't mean it's a terrible question. I've only ever had to use a Safari CSS hack once as a temporary solution and later got rid of it. I agree that you shouldn't have to target just Safari, but no harm in knowing how to do it. FYI, this hack only targets Safari 3, and also targets Opera 9.}'' This sample shows a positive attitude to a proposed question and makes some explanation. However, the amount of noisy information in such a long sample is so large that \approach ignores the positive sentiment conveyed in it and misclassifies it as neutral.
\end{itemize}

\section{Lessons Learned and Implications}\label{lesson}
% So far, we have answered the two research questions through experiments and demonstrated the effectiveness of \approach. 
In this section, we summarize the lessons learned from the investigation and try to propose some implications for future research:

$\bullet$ \noindent \textbf{Researchers can try to leverage the general sentimental expressions from the open domain, rather than unilaterally pursuing the domain-specific knowledge from the limited labeled data in SE.}

Previous studies found that sentiment analysis tools trained on non-technical texts are not adequate for SE tasks~\cite{IslamZ17,JongelingSDS17} and a lack of domain-specific knowledge is the main reason~\cite{IslamZ17}. Since then, many studies focused on how to use labeled SE-related texts to train SE-customized sentiment classifiers~\cite{IslamZ17,AhmedBIR17,ese/CalefatoLMN18}. However, in fact, there are many general sentimental expressions shared by SE and open domains. Therefore, we are interested whether the potential value of such information has been deeply explored in previous studies. On the other hand, the customization of sentiment analysis for SE is still at dawn, and thus the labeled data for SE are still too scarce to train a high-quality classification model. Therefore, we should not unilaterally place emphasis on learning the technical jargon based on existing already-labeled SE-related data. In this study, we demonstrate that learning the general sentimental information from large scale social media data can exactly improve sentiment analysis in SE. 

$\bullet$ \noindent \textbf{Emojis have potential values in SE, and it can play a role in more aspects in the future.} 

\approach is a promising attempt of employing emojis in SE tasks. In this study, we use emojis to tackle the problem of scarce labeled data that limits the performance of SE-customized sentiment analysis. Our experience can inform researchers to pay more attention to emojis as an instrument to investigate other sentiment-related topics in SE, such as activeness analysis~\cite{SinhaMG13}, emotion detection~\cite{msrMurgiaTAO14}, opinion mining~\cite{UddinK17a}, and politeness measurement~\cite{OrtuADTMT15}. For example, emotion detection can be quickly performed based on our study. This task aims to detect specific emotions of developers (e.g., joy, sadness, and anger). In previous studies, researchers performed it by manually labeling developer-generated posts in order to learn supervised classifiers~\cite{msrMurgiaTAO14}, but the labeled posts are limited. Since emojis are used to convey various emotions, \approach can be applied for such a task directly via the fine-tuning technique. 
% We can keep the first stage of \approach unchanged and only replace the $n$-dimension softmax layer in the second stage with an $m$-dimension one, where $m$ denotes the number of emotion classes. In this way, texts labeled with emotion labels can be used to train the emotion classifier in the second stage.

$\bullet$ \noindent \textbf{SEntiMoji can benefit the tasks that rely on identifying the sentiment of SE-related texts.} 

In a previous study, Lin \textit{et al.}~\cite{D0008ZBPLO18} argued that  no tool is ready for real usage of identifying sentiment expressed in SE-related discussions yet. However, from the results reported in Section~\ref{result}, we can see that \approach provides a possible solution to the sentiment-enabled tasks in SE.

Recall that the original intention of the Java Library dataset is to mine developers' opinions towards different software libraries, which can benefit the library recommendation task in the software development~\cite{D0008ZBPLO18}. In terms of the performance of SentiCR, nearly half of the sentences classified as positive or negative are actually false (precision@pos: 0.553, precision@neg: 0.546), which is far from being qualified for evaluating a software library. The precision levels of other existing methods are even lower. By comparison, \approach can achieve a comparable recall level but significantly improve  precision results over positive and negative samples (precision@pos: 0.849, precision@neg: 0.729). We inspect its confusion matrices on the Java Library dataset (see Table~\ref{confusion}) and find that the bad cases are mainly a result of classifying positive and negative samples as neutral rather than as the opposite sentiment polarity. It is promising for the aforementioned recommendation task. Although \approach could miss some sentimental posts, the software libraries predicted to be very positively or negatively discussed by \approach are credible due to its high precision in detecting sentimental expressions, which meets the criterion in such recommendation tasks that finding the right candidates is usually far more important than finding as many candidates as possible~\cite{WeiZYCFXRM17}. 

For other tasks such as sentiment detection for improving developers' productivity, \approach can also be potentially helpful. When a project manager finds that developers are suffering from negative sentiment reported by \approach, it is true with a high probability and worth paying attention to. Therefore, the manager can take necessary actions to defuse the situation. Additionally, \approach can benefit the empirical studies that investigate the correlation between developers' sentiments and other SE-related factors. For example, when exploring the correlation between negative sentiments and issue fixing time~\cite{OrtuADTMT15}, the conclusion might be biased if many detected negative posts are not really negative. The high precision of \approach on sentimental posts can alleviate this bias to some extent.

Finally, although \approach achieves satisfactory precision in comparison with previous methods, we should admit that more efforts should be devoted further to improving its recall level. If the recall level is further improved, \approach can be more qualified for the previously mentioned tasks in SE. For example, with a higher recall level, \approach can be more sensitive to negative emotions of developers and inform the project managers more promptly.

\begin{table}[!tp]
\small
\centering
\caption{Confusion matrices obtained by \approach on the Java Library dataset.}
\label{confusion}
\begin{tabular}{lrrrr}
\hline
& & \multicolumn{3}{c}{Predicted Polarity} \\
& & positive & neutral & negative \\ 
\hline
\multirow{3}{*}{\tabincell{l}{Ground\\Truth}} & positive & 43 & 84 & 4 \\
& neutral & 6 & 1,148 & 37 \\
& negative & 2 & 72 & 104 \\
\hline
\end{tabular}
\end{table}

$\bullet$ \noindent \textbf{The construction process of datasets can affect the evaluation of sentiment analysis methods, so the performance on different datasets should be analyzed more rationally.}

When comparing \approach with existing baseline methods, we find that performance gaps vary a lot on different datasets. It is not the first time that such a phenomenon is observed and reported. Novielli \textit{et al.}~\cite{NovielliGL08} attributed this result to the different labeling approaches (i.e., model-driven annotation or ad hoc annotation). In this study, we turn to the construction process of these datasets to seek the reason. For example, the Stack Overflow dataset is constructed based on the corpus filtered by SentiStrength. Therefore, when comparing the performance of SentiStrength and other methods on this dataset, we should be careful in order to avoid being biased. 

In addition, both the Stack Overflow and Code Review datasets are pre-processed to avoid imbalanced class distribution. It makes the classification task easier and may not match their target application scenarios. If researchers want to apply one sentiment analysis method to a technical Q\&A site such as Stack Overflow where the distribution of sentiments is highly imbalanced, they should not be blinded by the satisfactory results obtained on such datasets. Of course, we cannot deny or ignore that in some applications, the distribution of sentiments is relatively balanced. For these applications, evaluations on such processed datasets are meaningful. 
%What's more, these findings also remind future research to be careful when constructing and processing datasets. As casual sampling is considered to one of the top ten mistakes in data mining~\cite{nisbet2009handbook}, it is also highlighted in SE research~\cite{NagappanZB13}. To achieve general and reliable results, we should sample the data carefully in order to make it follow the distribution in real application scenarios.

%\noindent \textbf{(4) More studies about how software developers communicate should be encouraged.} 
%
%This paper is mainly inspired by studies about the non-verbal cues (e.g., emoticons and emojis) in SE platforms~\cite{ClaesMF18,githubemoji}. Such investigation about developers' communication can provide implications for the current natural language processing tasks in SE domain such as sentiment analysis and emotion detection. More specifically, the differences (not limited to the differences of sentimental expressions) between social media and communication in SE platforms should be highlighted, which can inspire the researchers how to leverage the current NLP methods proposed for social media data in the SE-specific tasks about developers' communication.

% \section{Limitations and Future Possibilities}\label{limitation}
% In this section, we first discuss some limitations including threats to construct validity, internal validity, and external validity.
\section{Threats to Validity}\label{limitation}
\textbf{Threats to construct validity} concern the relation between theory and observation. The JIRA dataset is originally labeled with different emotions. To use this dataset in our sentiment classification task, we follow previous studies to map the multi-class emotions to trinary sentiment polarity labels and filter some ambiguous samples. This process may violate the original distribution of this dataset and lower the difficulty of the classification task, which can affect the performance of different methods. However, luckily, we have four benchmark datasets for a comprehensive comparison and the superiority of \approach is not only observed on the JIRA dataset.

\textbf{Threats to internal validity} concern confounding factors that could affect the obtained results. In our study, they are mainly from the configuration of existing sentiment analysis methods. We replicate these methods by using their released pre-process scripts and recommended hyper-parameter settings. However, some hyper-parameters can be further tuned to improve the classification performance. In addition, different from GitHub and Stack Overflow posts whose code snippets are highlighted and can be easily removed, posts in the JIRA and Code Review datasets have mixed texts and code snippets. The scripts of some baseline methods such as SentiCR use manually defined rules such as searching for pre-defined programming keywords to identify and remove code snippets. However, simply removing keywords will violate the sentence structure, which may compromise the performance of our sentence-level \approach. Therefore, we do not adopt these naive rules to remove code snippets for \approach, which may affect its performance.

\textbf{Threats to external validity} concern the generalizability of our experimental results. Our evaluation has covered four mostly used benchmark datasets for sentiment analysis in SE. However, we still cannot claim that the performance of \approach can be generalized across datasets because the four datasets cannot represent all types of texts in SE. What's more, various sentiment analysis methods have been used in SE. In our study, we select only some representative ones for comparison.

\section{Conclusion}\label{conclusion}
In this study, we have proposed \approach, an emoji-powered learning approach for sentiment analysis in SE. This model is developed based on an existing representation model called DeepMoji. DeepMoji is pre-trained on Tweets and can represent texts with sentiment-aware vectors. As domain-specific knowledge is highlighted in the current SE-customized sentiment analysis, we also use GitHub posts to incorporate more technical jargon into DeepMoji. Then the fine-tuned representation model, as well as the manually labeled data, is used to train the final sentiment classifier. We evaluate the effectiveness of \approach on four benchmark datasets and \approach can significantly outperform the existing sentiment analysis methods in SE. Finally, we investigate the impact of the Tweets, GitHub posts, labeled data on the performance of \approach. The results demonstrate the importance of capturing the general sentimental expressions shared by technical and non-technical communication for sentiment analysis in SE.

We have released the data, code, trained models, and experiment results used in this study on \url{https://github.com/SEntiMoji/SEntiMoji}.
\section*{Acknowledgment}
This work was supported by the National Key R\&D Program of China under the grant number 2018YFB1004800, the National Natural Science Foundation of China under the grant number 61725201, and the Beijing Municipal Science and Technology Project under the grant number Z171100005117002.

\bibliographystyle{ACM-Reference-Format}
\bibliography{emojisentiment-bibliography}

\end{document}